\documentclass[12pt]{article}
\usepackage{graphicx}
\usepackage{psfig,epsfig}
\usepackage{caption}
\def\beq{\begin{equation}}
\def\eeq{\end{equation}}
\def\eighth{{\textstyle{1 \over 8}}} 
\def\half{{\textstyle{1 \over 2}}} 
\def\mnt{\hbox{$m_{\nu_\tau}$ }}
\def\nm{\hbox{$\nu_\mu$ }}
\def\nt{\hbox{$\nu_\tau$ }}
\def\quarter{{\textstyle{1 \over 4}}}

\let\vev\VEV

\begin{document}
\thispagestyle{empty}
\begin{titlepage}
\hfill hep-ph/9903474\\
\hspace*{1cm}\hfill FTUV/99-16\\
\hspace*{1cm}\hfill IFIC/99-17\\
\hspace*{1cm}\hfill UWThPh-1999-18\\
\vskip 0.3cm
\begin{center}
{\large \bf Top-Quark  Phenomenology in Models with
Bilinearly and Spontaneously Broken R-parity }
\vskip 0.5cm
{\large  Llu\'\i s Navarro$^1$},
{\large   Werner Porod$^2$},
{\large   and Jos\'e W.~F.~Valle$^1$}
\end{center}
\vskip 0.5cm
\begin{center}
{\sl
$^1$ Departament de F\'\i sica Te\`orica-IFIC, CSIC-Univ. de Val\`encia\\ 
Burjassot, Val\`encia 46100, Spain\\ 
http://neutrinos.uv.es\\
 $^2$Insitut f\"ur Theoretische Physik, Univ.~Wien \\
A-1090 Vienna, Austria
 }
\end{center}
\date{\today}
\vskip 2cm
\abstract{We study unconventional decays of the top-quark
in the framework of SUSY models with spontaneously broken R-parity.
In particular we discuss an effective theory which consists of the
MSSM plus bilinearly broken R-parity. We demonstrate that the decay
modes $t \to {\tilde \tau}^+_1 \, b$ and $t \to \tau^+ \, {\tilde b}_1
$ can have large branching ratios even in scenarios where the
tau-neutrino mass is very small. We show that existing Tevatron data
already probe the theoretical parameters, with promising prospects for
further improvement at the Run 2 of the Tevatron.}
%\pacs{ }
\end{titlepage}

\section{Introduction}

Although remarkably successful in the description of the phenomenology
of the strong and electroweak interactions, the Standard Model (SM)
leaves unanswered some issues such as the hierarchy problem and the
unification of the gauge couplings.  These have provided strong
impetus to the study of supersymmetric extensions \cite{HabKane}, in
particular those that break R-parity, RPV models, for short
\cite{old}. Of these we will focus on the case of bilinear R--Parity
Violation, BRPV for short~\cite{epsrad,BRPVtalk,BRPhiggs,others}.
They are well-motivated theoretically as they arise as effective
truncations of models where R--Parity is broken spontaneously
\cite{SRpSB} through right handed sneutrino vacuum expectation values
(VEV) $\vev{\tilde\nu^c_j}=v_{Rj} \neq0$.
These models open new possibilities for the study of the unification
of the Yukawa couplings~\cite{epsbtaunif}. In particular it has been
shown that in BRPV models bottom-tau unification may be
achieved at any value of $\tan\beta$. 
These models predict a plethora of novel processes \cite{BRPVtalkphen}
that could reveal the existence of SUSY in a totally different way,
not only through the usual missing momentum signature as predicted by
the Minimal Supersymmetric Standard Model (MSSM).  They provide a very
predictive approach to the violation of R--Parity, which renders the
systematic study of R-parity violating physics~\cite{BRPVtalkphen}
possible. Also, they are more restrictive than trilinear R--parity
violating (TRPV) models, especially in their supergravity formulation
with universal soft-breaking terms at the unification scale, as
in\cite{epsrad}.

Here we will consider the simplest superpotential which violates
R-Parity
\begin{equation} 
W_{R_p \hspace{-3.4mm} /} = W_{MSSM} 
+\epsilon_i \widehat L_i \widehat H_u \,,
\label{eq:Wsuppota}
\end{equation}
assuming that tri-linear terms are absent or suppressed, as would be
the case if their origin is gravitational~\cite{BJV}.  The
$\epsilon_i$ terms also violate lepton number in the $i$th generation
respectively. In models with spontaneously broken R--Parity
\cite{SRpSB} the $\epsilon_i$ parameters are then identified as equal
to some Yukawa coupling times $v_{Rj}$. This provides the main
theoretical motivation for adding explicitly BRPV to the MSSM
superpotential.

It has often been claimed that the BRPV term can be rotated away from
the superpotential by a suitable choice of the basis
\cite{HallSuzuki}.  If this were true the $\epsilon_i$ terms would be
unphysical.  However, one can show that, even though performing this
rotation of the superfields indeed eliminates the BRPV, RPV terms are
reintroduced in the form of TRPV.  Moreover, supersymmetry must be
broken and the presence of the $\epsilon_i$ terms in the superpotential
also introduces  R-parity violating terms $\varepsilon_{ab} (B_i
\epsilon_i L_i^a H^b_u)$ in the scalar sector, implying that the
vacuum expectation value $\vev{\tilde\nu_{li}}=v_i/\sqrt{2}$ is
non--zero. This in turn generates more R--parity and lepton number
violating terms inducing a tau-neutrino mass. Thus it is in general
impossible to rotate away the bilinear term in the Superpotential and
in the soft SUSY breaking potential at the same time.

In this model the top-quark gets additional decay modes, e.g. $t \to
{\tilde \tau}^+_1 \, b$.  We study these decays in view of the
Tevatron (top decays in TRPV models have been treated in
\cite{dreiner}) and show that existing Tevatron data already pose
restrictions on the parameter space.  This work is organized in the
following way: in Sect.~2 we discuss the model working out the
necessary details for the discussion of the top decays which will be
discussed in Sect.~3. In Sect.~4 we draw our conclusions.

\section{The model}

For simplicity we set from now on $\epsilon_1=\epsilon_2=0$, and in this
way, only tau--lepton number is violated. In this case, considering
only the third generation, the BRPV superpotential has the form:
\begin{equation} 
W_{R_p \hspace{-3.4mm} /}=\varepsilon_{ab}\left[
 h_t\widehat Q_3^a\widehat U_3\widehat H_u^b
+h_b\widehat Q_3^b\widehat D_3\widehat H_d^a
+h_{\tau}\widehat L_3^b\widehat R_3\widehat H_d^a
+\mu\widehat H_u^a \widehat H_d^b
+\epsilon_3\widehat L_3^a\widehat H_u^b\right]\,,
\label{eq:Wsuppot}
\end{equation}
where the first four terms correspond to the MSSM. The last term
violates tau--lepton number and therefore also R--Parity.  The soft SUSY
breaking potential is given by
\begin{equation}
V_{soft} = V_{soft,MSSM}
+B_2\epsilon_3\widetilde L_3 H_u+h.c
\end{equation}
where $V_{soft,MSSM}$ is the usual soft SUSY breaking potential of the
MSSM.

The scalar potential contains tadpoles
\begin{equation} 
V_{linear}=t_1^0\sigma^0_1+t_2^0\sigma^0_2+t_3^0\tilde\nu^R_{\tau}\,, 
\label{eq:Vlinear} 
\end{equation} 
where 
\begin{eqnarray} 
t_1^0&=&(m_{H_d}^2+\mu^2)v_d-B\mu v_u-\mu\epsilon_3v_3+ 
\eighth(g^2+g'^2)v_d(v_d^2-v_u^2+v_3^2)\,, 
\nonumber \\ 
t_2^0&=&(m_{H_u}^2+\mu^2+\epsilon_3^2)v_u-B\mu v_d+B_2\epsilon_3v_3- 
\eighth(g^2+g'^2)v_u(v_d^2-v_u^2+v_3^2)\,, 
\label{eq:tadpoles} \\ 
t_3^0&=&(m_{L_3}^2+\epsilon_3^2)v_3-\mu\epsilon_3v_d+B_2\epsilon_3v_u+ 
\eighth(g^2+g'^2)v_3(v_d^2-v_u^2+v_3^2)\,. 
\nonumber 
\end{eqnarray} 
and they are equal to zero at the minimum of the potential. Here $v_u$, $v_d$,
and $v_3$ are the VEVs of $H^0_u$, $H^0_d$, and ${\tilde \nu}_\tau$,
respectively.  $m_{H_u}$, $m_{H_d}$, and $m_{L_3}$ are the corresponding soft 
Susy-breaking mass parameters.

The charginos mix with the tau lepton. In a basis where  
$\psi^{+T}=(-i\lambda^+,\widetilde H_u^+,\tau_R^+)$ 
and $\psi^{-T}=(-i\lambda^-,\widetilde H_d^-,\tau_L^-)$, the charged 
fermion mass terms in the Lagrangian are 
\begin{equation} 
{\cal L}_m=-{1\over 2}(\psi^{+T},\psi^{-T}) 
\left(\matrix{{ 0} &  M_C^T \cr { M_C} &  
{ 0} }\right) 
\left(\matrix{\psi^+ \cr \psi^-}\right)+h.c. 
\label{eq:chFmterm} 
\end{equation} 
where the chargino/tau mass matrix is given by 
\begin{equation} 
{ M_C}=\left[\matrix{ 
M & {\textstyle{1\over{\sqrt{2}}}}gv_u & 0 \cr 
{\textstyle{1\over{\sqrt{2}}}}gv_d & \mu &  
-{\textstyle{1\over{\sqrt{2}}}}h_{\tau}v_3 \cr 
{\textstyle{1\over{\sqrt{2}}}}gv_3 & -\epsilon_3 & 
{\textstyle{1\over{\sqrt{2}}}}h_{\tau}v_d}\right] 
\label{eq:ChaM6x6} 
\end{equation} 
and $M$ is the $SU(2)$ gaugino soft mass.  Clearly, the chargino
sector decouples from the tau sector in the limit
$\epsilon_3=v_3=0$. As in the MSSM, the chargino mass matrix is
diagonalized by two rotation matrices $ U$ and $ V$ defined by
\beq
\chi_i^-=U_{ij}\, \psi_j^- \quad ; \quad \chi_i^+=V_{ij}\, \psi_j^+
\eeq
Then 
\beq
{U}^* {M_C} {V}^{-1} = {M_{CD}}
\eeq
where ${M_{CD}}$ is the diagonal charged fermion mass matrix.
The tau Yukawa coupling $h_{\tau}$ is chosen such that one of the  
eigenvalues is equal to the tau mass. This is calculated from the  
vacuum expectation values of the model through an exact tree 
level relation given by 
\begin{equation} 
h_{\tau}^2=\frac{2m_\tau ^2}{v_d^2}\left( \frac{% 
f+g(\varepsilon _3,v_3)}{f-\frac 2{v_d^2}h(\varepsilon _3,v_3)}\right)   
\label{eq:htauf} 
\end{equation} 
The functions $f$, $g$ and $h$ are given in ref.~\cite{chargedhiggs}.

In our model, the tau neutrino acquires mass due to a mixing between
the neutralino sector and the tau--neutrino.  In the basis $\psi^{0T}=
(-i\lambda',-i\lambda^3,\widetilde{H}_d^1,\widetilde{H}_u^2,\nu_{\tau})$
the neutral fermions mass terms in the Lagrangian are given by
\begin{equation} 
{\cal L}_m=-\frac 12(\psi^0)^T{ M}_N\psi^0+h.c.   
\label{eq:NeuMLag} 
\end{equation} 
where 
\begin{equation} 
{ M}_N=\left[  
\begin{array}{ccccc}  
M^{\prime } & 0 & -\frac 12g^{\prime }v_d & \frac 12g^{\prime }v_u & -\frac  
12g^{\prime }v_3 \\   
0 & M & \frac 12gv_d & -\frac 12gv_u & \frac 12gv_3 \\   
-\frac 12g^{\prime }v_d & \frac 12gv_d & 0 & -\mu  & 0 \\   
\frac 12g^{\prime }v_u & -\frac 12gv_u & -\mu  & 0 & \epsilon _3 \\   
-\frac 12g^{\prime }v_3 & \frac 12gv_3 & 0 & \epsilon _3 & 0  
\end{array}  
\right] 
\label{eq:NeuM5x5} 
\end{equation} 
and $M'$ is the $U(1)$ gaugino soft mass. This neutralino/neutrino mass  
matrix is diagonalized by a $5\times 5$ rotation matrix $ N$ such that 
\begin{equation} 
{ N}^*{ M}_N{ N}^{-1}={\rm diag}(m_{\chi^0_1},m_{\chi^0_2}, 
m_{\chi^0_3},m_{\chi^0_4},m_{\nu_{\tau}}) 
\label{eq:NeuMdiag} 
\end{equation} 
% 
%where by definition the eigenstate $F_5^0$ is the neutrino--tau, i.e., 
%with the largest tau component $(N_{i5})^2$. For future reference we
%note that
The physical states $\chi_j^0$ are given by:
\beq
\psi_j^0={N^*}_{kj}\, \chi_j^0 \,.
\eeq 

There is also a mixing between the charged Higgs boson and the staus.
The mass matrix of the charged scalars is given by:
\begin{equation} 
L_m=-[H_d^-,H_u^-,\tilde\tau_L^-,\tilde\tau_R^-] 
{M_{S^{\pm}}^2}\left[\matrix{H_d^+ \cr H_u^+ \cr \tilde\tau_L^+ \cr 
\tilde\tau_R^+}\right]+ h.c. \, . 
\label{eq:Vquadratic} 
\end{equation} 
For convenience we divide this $4\times4$ matrix into 
$2\times2$ blocks in the following way: 
\begin{equation} 
{M_{S^{\pm}}^2}=\left[\matrix{ 
{ M_{HH}^2} & { M_{H\tilde\tau}^{2T}} \cr 
{ M_{H\tilde\tau}^2} & { M_{\tilde\tau\tilde\tau}^2} 
}\right]  \, .
\label{eq:subdivM} 
\end{equation} 
The charged Higgs block is given by 
\begin{eqnarray} 
&& { M_{HH}^2}= 
\label{eq:subMHH} \\ \nonumber \\ 
&& \!\!\!\!\!\!\left[\matrix{ 
B\mu{{v_u}\over{v_d}}+\quarter g^2(v_u^2-v_3^2)+\mu\epsilon_3 
{{v_3}\over{v_d}}+\half h_{\tau}^2v_3^2+{{t_1}\over{v_d}} 
& B\mu+\quarter g^2v_dv_u 
\cr B\mu+\quarter g^2v_dv_u 
& B\mu{{v_d}\over{v_u}}+\quarter g^2(v_d^2+v_3^2)-B_2\epsilon_3 
{{v_3}\over{v_u}}+{{t_2}\over{v_u}} 
}\right] 
\nonumber 
\end{eqnarray} 
and the stau block is given by 
\begin{eqnarray} 
&& { M_{\tilde\tau\tilde\tau}^2}= 
\label{eq:subtautau} \\ \nonumber \\ 
&& \!\!\!\!\!\!\!\!\!\!\!\!\!\!\left[\matrix{ 
\half h_{\tau}^2v_d^2-\quarter g^2(v_d^2-v_u^2)+\mu\epsilon_3 
{{v_d}\over{v_3}}-B_2\epsilon_3{{v_u}\over{v_3}}+{{t_3}\over{v_3}} 
& {1\over{\sqrt{2}}}h_{\tau}(A_{\tau}v_d-\mu v_u) 
\cr {1\over{\sqrt{2}}}h_{\tau}(A_{\tau}v_d-\mu v_u) 
& m_{E_3}^2+\half h_{\tau}^2(v_d^2+v_3^2) 
-\quarter g'^2(v_d^2-v_u^2+v_3^2) 
}\right] 
\nonumber 
\end{eqnarray} 
We recover the usual stau and Higgs--mass matrices in the limit
$v_3=\epsilon_3=0$ (we need to replace the expression of the third
tadpole in Eq.~(\ref{eq:tadpoles}) before taking the limit).  The
mixing between the charged Higgs sector and the stau sector is given
by the following $2\times2$ block:
\begin{equation} 
{ M_{H\tilde\tau}^2}=\left[\matrix{ 
-\mu\epsilon_3-\half h_{\tau}^2v_dv_3+\quarter g^2v_dv_3 
& -B_2\epsilon_3+\quarter g^2v_uv_3 
\cr -{1\over{\sqrt{2}}}h_{\tau}(\epsilon_3v_u+A_{\tau}v_3) 
& -{1\over{\sqrt{2}}}h_{\tau}(\mu v_3+\epsilon_3v_d) 
}\right] 
\label{eq:subHtau} 
\end{equation} 
and as expected, this mixing vanishes in the limit $v_3=\epsilon_3=0$.
The charged scalar mass matrix in Eq.~(\ref{eq:subdivM}), after
setting $t_1=t_2=t_3=0$, has determinant equal to zero, since one of
the eigenvectors corresponds to the charged Goldstone boson ``eaten''
by the W boson.  The mass matrices in Eqs.~(\ref{eq:subdivM}) are
diagonalized by a rotation matrix $R_{S^{\pm}}$:
\begin{equation} 
\left[\matrix{G^+ \cr H^+ \cr \tilde\tau_1^+ \cr \tilde\tau_2^+}\right]= 
{ R_{S^{\pm}}}\left[\matrix{ 
H_d^+ \cr H_u^+ \cr \tilde\tau_L^+ \cr \tilde\tau_R^+}\right] .
\label{eq:eigenvectors} 
\end{equation} 
and the eigenvalues are
$\rm{diag}(0,m_{H^{\pm}}^2,m_{\tilde\tau_1^{\pm}}^2, 
m_{\tilde\tau_2^{\pm}}^2)={ R_{S^{\pm}}}{ M_{S^{\pm}}^2} 
{ R_{S^{\pm}}^T}$.

A similar mixing occur between the neutral Higgs bosons and the real
part of the tau-sneutrino and between the pseudoscalar Higgs and the
imaginary part of the tau-sneutrino \cite{epsrad,BRPhiggs}. We denote
the resulting scalar (pseudoscalar) state by $S^0_i$ ($P^0_i$) with
$m_{S^0_i} < m_{S^0_j}$ for $i<j$.

\section{R-parity Violating Top Decays}

\begin{figure}
\includegraphics[height=11cm,width=13cm]{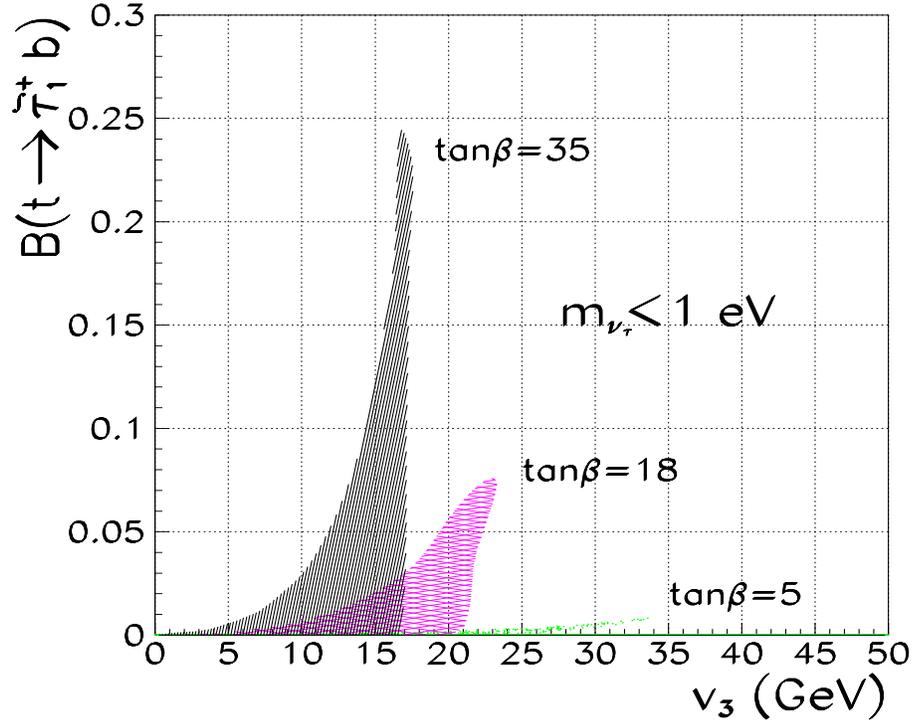}
\includegraphics[height=11cm,width=13cm]{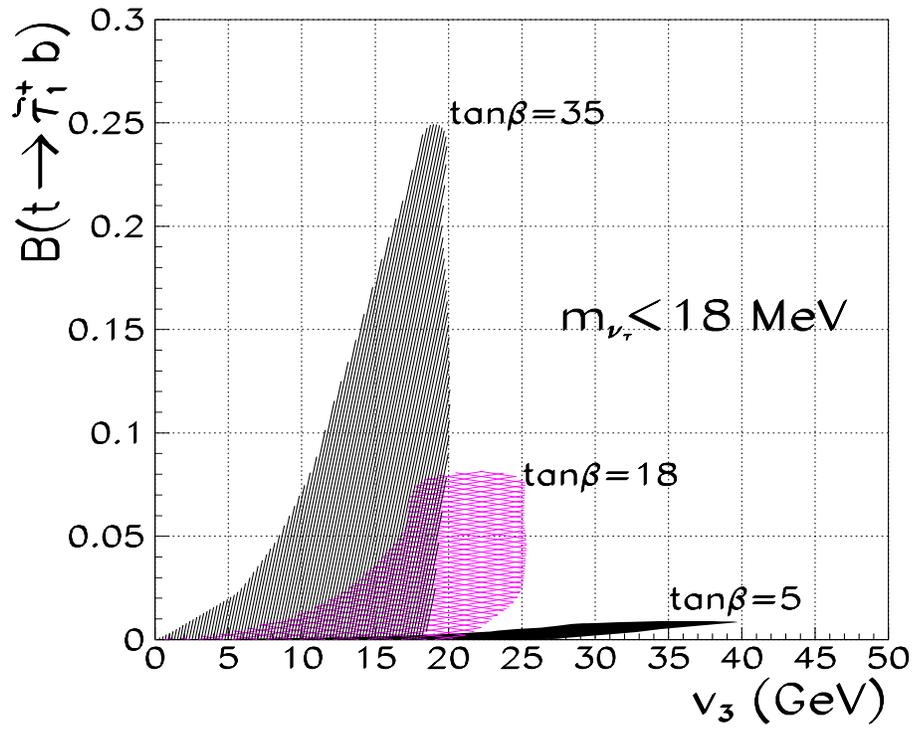}
%\setlength{\unitlength}{1mm}
%\begin{picture}(150,215)
%\put(-2,93){\mbox{\epsfig{figure=BrTopStau_MeV.eps,height=14.0cm,width=15cm}}}
%\put(-2,220){\makebox(0,0)[bl]{{\Large \bf a) }}}
%\put(0,209){\makebox(0,0)[bl]{{\Large
%             $\bf \mbox{BR}(t \to {\tilde \tau}_1 \, b)$}}}
%\put(148.5,115){\makebox(0,0)[br]{{\Large \bf $\bf v_3$~[GeV]}}}
%\put(55,195){\makebox(0,0)[br]{{\large \bf $m_{\nu_\tau}<18$~MeV}}}
%\put(98,192){\makebox(0,0)[br]{{\large \bf $\tan \beta = 35$}}}
%\put(112,161){\makebox(0,0)[br]{{\large \bf $\tan \beta = 18$}}}
%\put(141,141){\makebox(0,0)[br]{{\large \bf $\tan \beta = 5$}}}
%\put(-2,-22){\mbox{\epsfig{figure=BrTopStau_eV.eps,height=14.0cm,width=15cm}}}
%\put(-2,105){\makebox(0,0)[bl]{{\Large \bf b) }}}
%\put(0,94){\makebox(0,0)[bl]{{\Large
%             $\bf \mbox{BR}(t \to {\tilde \tau}_1 \, b)$}}}
%\put(148.5,0){\makebox(0,0)[br]{{\Large \bf $\bf v_3$~[GeV]}}}
%\put(49,79){\makebox(0,0)[br]{{\large \bf $m_{\nu_\tau}<1$~eV}}}
%\put(85,65){\makebox(0,0)[br]{{\large \bf $\tan \beta = 35$}}}
%\put(102,47){\makebox(0,0)[br]{{\large \bf $\tan \beta = 18$}}}
%\put(141,26){\makebox(0,0)[br]{{\large \bf $\tan \beta = 5$}}}
%\end{picture}
\caption[]{Branching Ratio BR($t \to {\tilde \tau}_1^+ \, b)$ as a function
of $v_3$ for different values of $\tan \beta$. The other parameters are given
in the text. We have imposed the constraint $\mnt < 18$~MeV {\bf (a)} and
$\mnt < 1$~eV {\bf (b)}.}
\label{rpvtoptanbeta}
\end{figure}

One of the major successes of Tevatron has been the discovery of the
top-quark \cite{Topdiscovery}. The large top mass implies a relatively
small production cross section at the Tevatron. As a result the sum of
all branching ratios of the top decays except $t \to W^+ \, b$ is only
restricted to be smaller than approximately 25 \% \cite{TopEx}.
In the MSSM the top can decay according to: $t \to W^+ \, b$, $t \to
H^+ \, b$, $t \to {\tilde \chi}^0_1 \, {\tilde t}_1$, and 
$t \to {\tilde\chi}^+_1 \, {\tilde b}_1 $
(for their discussion in the MSSM see e.g.~\cite{topMSSM1,topMSSM2} and
references therein).
The last mode is only listed for completeness, because it is
practically ruled out by existing chargino and squark searches at LEP2
\cite{LEP2susy}.  In the BRPV model additional decay modes occur:
\begin{eqnarray}
 t &\to& {\tilde \tau}^+_1 \, b \, , \\
 t &\to& \tau^+ \, {\tilde b}_1 \, , \\
 t &\to& \nu_\tau \, {\tilde t}_1 \, .
\end{eqnarray}
For the following discussion we have randomly chosen {\cal O}($10^4$)
points in the SUSY parameter space imposing the relevant experimental
constraints on SUSY searches.
The MSSM bounds on sparticles are in general not directly applicable
to broken R-parity models and a reanalysis of the LEP and Tevatron
data is necessary in order to determine the corresponding bounds. In
the particular case of charginos such a reanalysis has been made and
it was found that the bound on the chargino mass in the BRPV model is
essentially the same as in the MSSM \cite{tauJ}. For definiteness we
have taken: $m_{ {\tilde t}_1}, m_{ {\tilde b}_1} > 80$~GeV,
$m_{S^0_1} > 70$~GeV, $min(m_{H^+}, m_{ {\tilde \tau}_1}) > 70$~GeV,
and $m_{{\tilde \chi}^+_1} > 90$~GeV.  Moreover, we have imposed the
\nt mass constraint $m_{\nu_\tau} < 18$~MeV \cite{mnutau97}.  Here,
one should bear in mind that the heavy \nt possibility is certainly
allowed by the present Super Kamiokande data, as they may be accounted
for either by conversions of \nm to sterile neutrinos
\cite{atmsterile}, flavour changing muon-neutrino interactions 
\cite{atmFC} or \nm decay ~\cite{atmdecay}. On the other
hand cosmological and astrophysical limits on \mnt are obviated in the
presence of neutrino decay and annihilation channels involving
majorons, present in the model with spontaneous breaking. For a review
see ref.~\cite{fae} and for recent references see e.g.~\cite{BBN}.
Therefore, this is the only conservative limit one can really apply on
the tau-neutrino mass.  However, as we will see in the following
discussion of Fig.~\ref{rpvtoptanbeta}, the R-parity violating
branching ratios may be sizeable even with a small tau-neutrino mass.

As an illustrative example we show in Fig.~\ref{rpvtoptanbeta} the
branching ratio for $t \to {\tilde \tau}^+_1 \, b$ as a function of
the R-parity violating parameter $v_3$ for different values of $\tan
\beta = v_2 / v_1$. We have varied $-500$~GeV $< \epsilon_3 < 0$~GeV
and $0$~GeV $< B_2 < 500$~GeV. The remaining parameters have been
fixed as: $M = 180$~GeV, $\mu = 200$~GeV, $M_{E_3} = 285$~GeV, $A_\tau
= 280$~GeV, $M_Q = 285$~GeV, $M_U = 180$~GeV, $M_D = 190$~GeV, $A_t =
320$~GeV, $A_b = 120$~GeV, and $B = 50$~GeV.  We have imposed the
constraint $\mnt < 18$~MeV {\bf (a)} and $\mnt < 1$~eV {\bf (b)}. One
immediately observes the following: (i) The larger $\tan \beta$ the
larger is the branching ratio. (ii) The bound on $\mnt$ does not lead
to smaller branching ratios but to smaller allowed values of $v_3$ for
fixed $\epsilon_3$.  The first fact is understood in the following
way: The stau mixes mainly with $H_d$ (since $\tilde \tau_L$ and $H_d$
have the same gauge quantum numbers).  There is a strong mixing
between the stau and the charged Higgs for the points where the
branching ratio is above $\sim 10\%$.  Moreover, the coupling $H_d \,
t \, b$ is proportional to $h_b$ which grows with $\tan \beta$.  For
the second point note that $\mnt$ is proportional to $\epsilon_3 v_1 +
\mu v_3$ \cite{BRPVtalk}. Moreover, one can keep in principal $\mnt$
fixed and one is still able to change the partial width $\Gamma(t \to
{\tilde \tau}^+_1 \, b)$ by varying $B_2$ and therefore the
Higgs--Stau mixing.
\begin{figure}
\setlength{\unitlength}{1mm}
\begin{picture}(150,215)
\put(-2,92){\mbox{\epsfig{figure=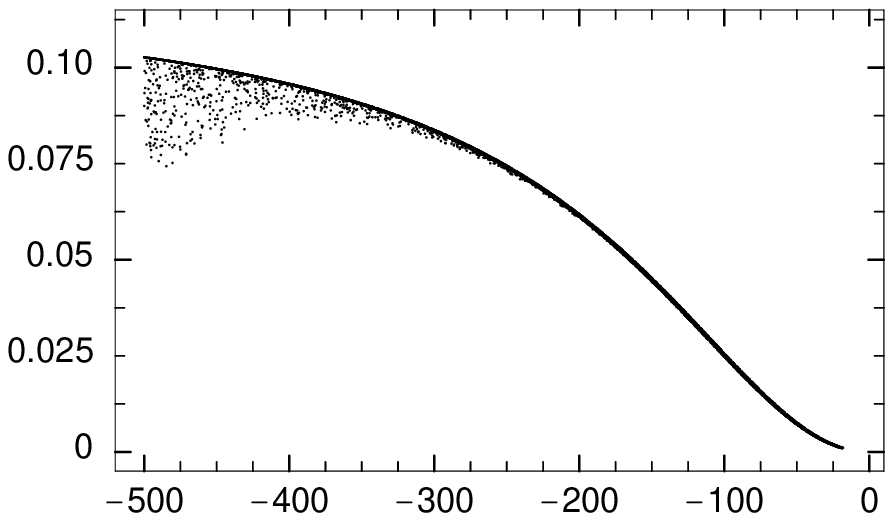,
                          height=14.5cm,width=15cm}}}
\put(-2,220){\makebox(0,0)[bl]{{\Large \bf a) }}}
\put(0,209){\makebox(0,0)[bl]{{\Large
             $\bf \mbox{BR}(t \to {\tilde b}_1 \, \tau)$}}}
\put(148.5,117){\makebox(0,0)[br]{{\Large \bf $\bf \epsilon_3$~[GeV]}}}
\put(-2,-23){\mbox{\epsfig{figure=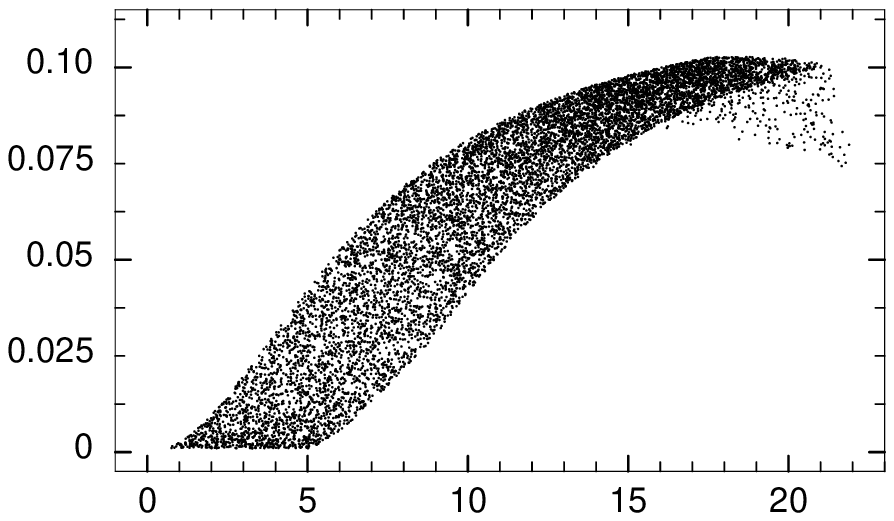,
                                  height=14.5cm,width=15cm}}}
\put(-2,105){\makebox(0,0)[bl]{{\Large \bf b) }}}
\put(0,94){\makebox(0,0)[bl]{{\Large
             $\bf \mbox{BR}(t \to {\tilde b}_1 \, \tau)$}}}
\put(148.5,2){\makebox(0,0)[br]{{\Large \bf $\bf v_3$~[GeV]}}}
\end{picture}
\caption[]{Branching Ratios for $t \to {\tilde b}_1 \, \tau^+$ as a
function of the R-parity violating parameters $\epsilon_3$ {\bf (a)} and
$v_3$ {\bf (b)}. Here we have assumed $\mnt<18$~MeV and the other parameters 
are given in the text.}
\label{topsbot}
\end{figure}

In Fig.~\ref{topsbot} we show the branching ratios 
$\mathrm{BR}(t \to \tau^+ \, {\tilde b}_1)$ as a function of $\epsilon_3$ 
and $v_3$ 
for $\tan \beta = 35$ and the other parameters as above (in the first case
we have taken $0< v_3 < 25$~GeV). 
$m_{{\tilde b}_1}$ grows with decreasing  $\tan \beta$ if one keeps the other 
parameters fixed and therefore this decay mode will be kinematically forbidden
for small $\tan \beta$.
In Fig.~\ref{topsbot}a we see a
strong correlation between $\epsilon_3$ and 
$\mathrm{BR}(t \to \tau^+ \, {\tilde b}_1)$. This can be understood in the 
following way: in the
chargino mass matrix (Eq.~(\ref{eq:ChaM6x6})) the mixing between the
leptons and the charginos disappears if one does the following rotation
of the superfields: $\widehat H_d \to N (\mu \widehat H_d - \epsilon_3
\widehat L_3)$ and $\widehat L_3 \to N (\mu \widehat L_3 + \epsilon_3
\widehat H_d)$ (N being the normalization). In this basis the coupling
between $t$, $\tau$, and ${\tilde b}_1$ is proportional $N \, h_b
\,\epsilon_3$ leading to this feature.  In Fig.~\ref{topsbot}b we
show the same branching ratio as a function of $v_3$. The fact that
the branching ratio is not larger than $\sim10\%$ is a consequence of
the limit on $m_{\nu_\tau}$. The parameters $v_3$ and $\epsilon_3$ are
responsible for the mixing between the charginos and the
tau-lepton. This mixing is the reason for the band observed in the
figure.

\begin{figure}
\setlength{\unitlength}{1mm}
\begin{picture}(150,215)
\put(-2,91){\mbox{\epsfig{figure=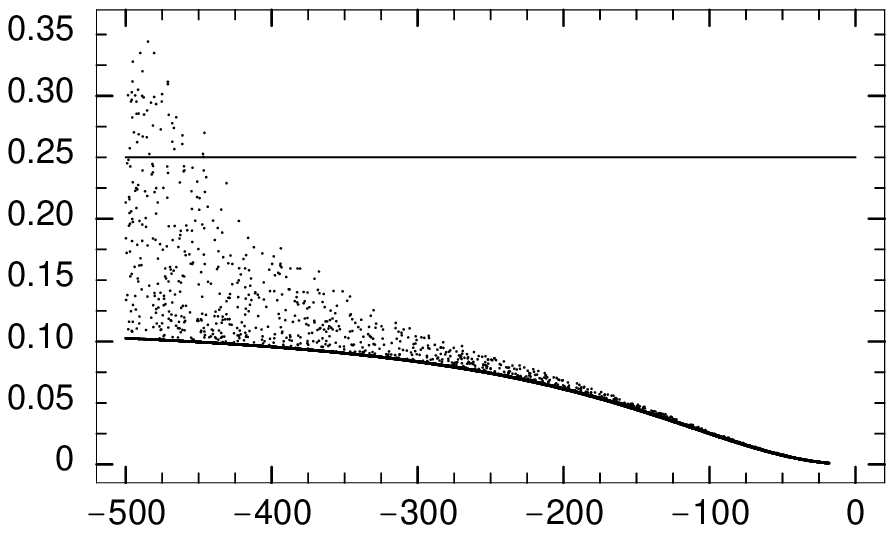,
                          height=14.5cm,width=15cm}}}
\put(-2,220){\makebox(0,0)[bl]{{\Large \bf a) }}}
\put(0,210){\makebox(0,0)[bl]{{\Large
             $\bf 1-\mbox{BR}(t \to  W \, b)$}}}
\put(52,185){\makebox(0,0)[bl]{{\large \bf Experimental Bound }}}
\put(148.5,116){\makebox(0,0)[br]{{\Large \bf $\bf \epsilon_3$~[GeV]}}}
\put(-2,-24){\mbox{\epsfig{figure=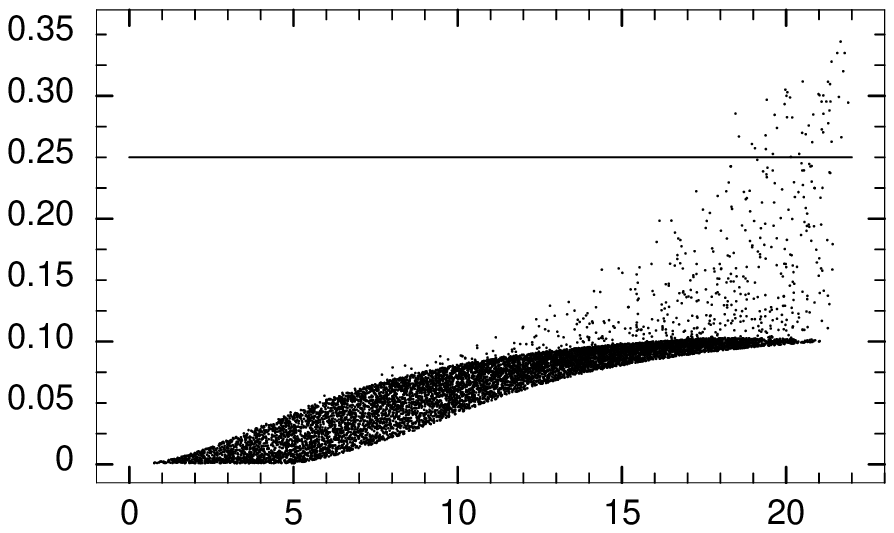,
                                  height=14.5cm,width=15cm}}}
\put(-2,105){\makebox(0,0)[bl]{{\Large \bf b) }}}
\put(0,95){\makebox(0,0)[bl]{{\Large
             $\bf 1-\mbox{BR}(t \to W \, b)$}}}
\put(32,70){\makebox(0,0)[bl]{{\large \bf Experimental Bound }}}
\put(148.5,1){\makebox(0,0)[br]{{\Large \bf $\bf v_3$~[GeV]}}}
\end{picture}
\caption[]{Sum of all branching Ratios different from $t \to W~b $ as a
function of $\epsilon_3$ {\bf (a)} and $v_3$ {\bf (b)}. The parameters are
chosen as in Fig.~\ref{topsbot}.}
\label{topdecaystot}
\end{figure}

In Fig.~\ref{topdecaystot} 
we show the sum of all branching ratios for non-standard decay modes
of the top quark as a function of $\epsilon_3$ (a) and $v_3$ (b).  
One can see that existing Tevatron data already exclude
parts of the parameter space which is not excluded by other data.
 In both cases we have cascade decays:\\
\begin{tabular}{lll}
$t \to {\tilde \tau}^+_1 \, b$ & $\to \tau^+ \, \nu_\tau \, b$ & \\
           & $\to \tau^+ \, {\tilde \chi}^0_1 \, b$ &
                      $\to \tau^+ \, f \, \bar{f} \, \nu_\tau \, b$ \\
           & &   $\to \tau^+ \, f \, \bar{f}' \, \tau^\pm \, b$ \\
           & $\to \nu_\tau \, {\tilde \chi}^+_1 \, b$ &
                      $\to \nu_\tau \, f \, \bar{f}' \, \nu_\tau \, b$ \\
           & &   $\to \nu_\tau \, f \, \bar{f} \, \tau^+ \, b$ \\
           &  $\to c \, s \, b$ & \\
$t \to \tau^+ \, {\tilde b}_1$ & $\to \tau^+ \, \nu_\tau \, b$ & \\
           & $\to \tau^+ \, {\tilde \chi}^0_1 \, b$ &
                      $\to \tau^+ \, f \, \bar{f} \, \nu_\tau \, b$ \\
           & &   $\to \tau^+ \, f \, \bar{f}' \, \tau^\pm \, b$ 
\end{tabular} \\

In most cases there are two $\tau$--leptons and two $b$-quarks in the
final state plus the possibility of additional leptons and/or jets.
Therefore, $b$-tagging and a good $\tau$ identification are important
for extracting these final states. Moreover, there is in general a
large multiplicity of charged particles in the final state which
should be helpful in reducing the background. The background stems
mainly from the production of one or two gauge bosons plus additional
jets.
The conclusion in similar cases \cite{LeCompte98} has been that within
its next run Tevatron will be sensitive for branching ratio
measurements up to $10^{-3} - 10^{-2}$ depending on the
mode. Therefore, the observation of one the additional decay modes at
the run 2 of Tevatron will give a strong hint on the underlying
R-parity violating parameters.

We have performed a similar scan for small $\tan \beta$ ($\le 5$). Here we have
found that the non-standard decay modes are smaller:~$1-\mathrm{BR}(t\to
W~b)<0.02$. This happens because their decay widths are in both cases
proportional to the bottom Yukawa coupling squared.
In case of $t \to {\tilde \tau}^+_1 \, b$ this can be already seen from 
Fig.~\ref{rpvtoptanbeta} and in case of 
$t \to \tau^+ \, {\tilde b}_1$ is clear from the discussion of
Fig.~\ref{topsbot}.
Up to now we have not discussed the decay 
$t \to \nu_\tau \, {\tilde t}_1$ because it turns out that the branching
ratio for this mode is tiny in the parameter region studied. The reason
is that the Higgsino ${\tilde H}_u$ hardly mixes with $\nu_\tau$. Moreover,
the coupling is proportional to $\mnt / m_W$.

\section{Conclusions}

We have seen that the R-parity violating decay modes of the top
quark $t \to {\tilde \tau}^+_1 \, b$ and 
$t \to {\tilde b}_1 \,  \tau^+$ can have large branching ratios, 
especially if the tau neutrino
is heavy.  We have also shown that existing Tevatron data already
probe the theoretical parameters, and the prospects for further
improvement at the Run 2 of the Tevatron are promising.
Finally we have also verified that the magnitude of the R-parity
violating top decay branching ratios considered here remains large
even when the \nt mass becomes very small, as perhaps indicated by the
recent atmospheric neutrino data.

\section*{Acknowledgements} 

This work was supported by DGICYT under grant number PB95-1077, by the
TMR network grant ERBFMRXCT960090 the European Union and by Accion
Integrada Hispano-Austriaca. L.~N. was supported by CSIC.
W.~P.~was supported by the ''Fonds zur F\"orderung der wissenschaftlichen
Forschung'' of Austria, project No. P13139-PHY.

\end{document}